\documentclass[graybox]{svmult}

\usepackage{type1cm}        
\usepackage{makeidx}         
\usepackage{graphicx}        
\usepackage{multicol}        
\usepackage[bottom]{footmisc}

\usepackage{newtxtext}       %
\usepackage{newtxmath}       
\usepackage[utf8]{inputenc}
\usepackage{amsmath}
\usepackage{subcaption}
\usepackage{caption}
\usepackage{xcolor}
\usepackage{soul}
\usepackage{multirow}

\usepackage{url}
\usepackage{physics}
\definecolor{lars}{rgb}{0.4, 0.0, 0.6}

\definecolor{GL}{rgb}{0.8, 0.4, 0.8}

\definecolor{ste}{rgb}{0., 0.26, 0.15}

\definecolor{cadmiumred}{rgb}{0.89, 0.0, 0.13}

\definecolor{max}{rgb}{0.9, 0.0, 0.6}

\makeindex             

\begin{document}

\title*{Comparison of outlier detection methods on astronomical image data}
\titlerunning{Comparison of Outlier Detection Methods on SDSS Image Data}

\author{Lars Doorenbos, Stefano Cavuoti, Massimo Brescia, Antonio D'Isanto, Giuseppe Longo} 
\authorrunning{Doorenbos et al. 2020}

\institute{Lars Doorenbos \at Bernoulli Institute for Mathematics, Computer Science and Artificial Intelligence, University of Groningen,  Nijenborgh 9, 9747AG Groningen, The Netherlands \email{larsdoorenbos@msn.com}
\and Stefano Cavuoti \at Department of Physics, University of Naples  Federico II, Strada Vicinale Cupa Cintia, 21, I-80126 Napoli, Italy. \at INAF - Astronomical Observatory of Capodimonte, Salita Moiariello 16, I-80131 Napoli, Italy \email{stefano.cavuoti@gmail.com} 
\and M. Brescia \at INAF - Astronomical Observatory of Capodimonte, Salita Moiariello 16, I-80131 Napoli, Italy. \email{massimo.brescia@inaf.it} 
\and  Antonio D'Isanto \at Astroinformatics group, Heidelberg Institute for Theoretical Studies, Schloss-Wolfsbrunnenweg 35, 69118 Heidelberg, Germany \email{antonio.disanto@h-its.org}
\and Giuseppe Longo \at Department of Physics, University of Naples  Federico II, Strada Vicinale Cupa Cintia, 21, I-80126 Napoli, Italy. \email{longouniversita@gmail.com}  }
%
\maketitle

\textit{Preprint version of the manuscript to appear in the Volume ``Intelligent Astrophysics'' of the series ``Emergence, Complexity and Computation'', Book eds. I. Zelinka, D. Baron, M. Brescia, Springer Nature Switzerland, ISSN: 2194-7287}\newline

\abstract*{Among the many challenges posed by the huge data volumes produced by the new generation of astronomical instruments there is also the search for rare and peculiar objects. Unsupervised outlier detection algorithms may provide a viable solution. In this work we compare the performances of six methods: the Local Outlier Factor, Isolation Forest, k-means clustering, a measure of novelty, and both a normal and a convolutional autoencoder. These methods were applied to data extracted from SDSS stripe 82. After discussing the sensitivity of each method to its own set of hyperparameters, we combine the results from each method to rank the objects and produce a final list of outliers.}

\abstract{Among the many challenges posed by the huge data volumes produced by the new generation of astronomical instruments there is also the search for \textit{rare} and \textit{peculiar} objects. Unsupervised outlier detection algorithms may provide a viable solution. In this work we compare the performances of six methods: the Local Outlier Factor, Isolation Forest, k-means clustering, a measure of novelty, and both a normal and a convolutional autoencoder. These methods were applied to data extracted from SDSS stripe 82. After discussing the sensitivity of each method to its own set of hyperparameters, we combine the results from each method to rank the objects and produce a final list of outliers.}

\section{Introduction}

As it has been amply discussed in the literature (cf. \cite{Harwit_1,Harwit_CD}) the discovery of most new astronomical objects and phenomena can be interpreted in terms of either an enlargement or better sampling of the observable parameter space (OPS) or in terms of an increased capability to extract patterns and trends. An example of the first type  being the discovery of quasars (due to the opening of a new dimension defined by the radio fluxes), while an example of the second type can be the discovery of the fundamental plane of elliptical galaxies (\cite{1987ApJ...313...59D}). So far, our capability to explore the OPS has been strongly limited by the human factor, i.e. by the difficulties encountered by the human brain in finding patterns or outliers in OPS with more than three dimensions.\\
The OPS defined by modern panchromatic multi-epoch surveys has a very large amount of dimensions, with more than several hundreds of parameters measured for each object. While on the one hand it poses large computational problems, on the other it offers unprecedented possibilities in moving from the so called ``serendipitous discoveries'' to a more systematic search for the \textit{rare} and of the \textit{unknown}.

This aspect, by considering also the huge amounts of data collected by modern sky surveys, implies that it is no longer feasible to go through all observations by hand to retrieve and flag the interesting objects. 
The unbiased and automatic data-driven capability of methods based on the paradigms of Machine Learning (ML), to explore the OPS and to extract hidden correlations among data, may achieve a realistic possibility to perform a manageable selection of peculiar objects, most likely to be interesting.
In astrophysical terms, the peculiar objects, or outliers, are mostly characterized by their intrinsic singularity within a data distribution. Satellites, imaging artifacts, interacting galaxies, lensed objects or other unexpected entities are just a few of examples of outliers that could emerge from a distribution. ML methods may efficiently analyze a dataset and isolate such peculiar candidates, by proceeding in two different ways: (i) supervised, i.e. by inferring knowledge of the nature of objects from a limited number of samples during training and by generalizing the obtained insights in order to recognize outliers in new data; (ii) unsupervised, i.e. by self-organizing their internal structure, just driven by data themselves, in order to recognize commonalities among data and isolating potential outliers as single entities within the parameter space. 
The supervised approach has an important downside, since it limits the capability of a method to recognize only expected (hence known) peculiarities, i.e. those provided within the training set. Therefore, we preferred to proceed in an unsupervised way and to perform a totally unbiased data exploration.\\
Among the varieties of unsupervised methods available, by considering the known heuristic process at the base of any ML approach, we selected a set of models and performed a comparison of their performance in extracting outliers. These methods were: (a) Local Outlier Factor \cite{breunig2000lof}, (b) Isolation Forest \cite{liu2008isolation}, (c) k-means \cite{lloyd1982least}, (d) a recently introduced novelty measure \cite{hajer2018novelty} (hereinafter named modified novelty or MN), and (e) both a normal and a convolutional autoencoder \cite{lyudchik2016outlier,ribeiro2018study}.\\
The comparison was driven not only in terms of the efficiency and accuracy to extract peculiar objects, but also regarding the capability to recognize common objects as potential outliers, trying to investigate the  perfomance in case of decision discrepancies.\\
In this work these methods were used to detect the most abnormal points in the training dataset. These can be singular observations, or small clusters of relatively normal data, lying far away from the normal dataset distribution within the feature space. Both  types were considered as outliers. 
Alternatively, 
one could consider the training dataset as ``normal'' and use the trained models for the detection of anomalies in a different test set by searching for the largest deviations from the observations from the training set \cite{pimentel2014review}. This approach will be explored in a forthcoming project. \\ 
Most methods suggested in the literature for the detection of outliers in sky surveys, focus on spectroscopic or tabular data. Examples include Chaudhary et al. \cite{chaudhary2002very}, who proposed a new outlier detection method, using five-dimensional SDSS data. In Fustes et al. \cite{fustes2013approach} self-organizing maps were used to find outlying spectra. Giles et al. \cite{giles2018systematic} used a variant of the DBSCAN clustering algorithm to detect outliers in derived light curve features.
The work more pertinent to that presented here is the one proposed by Baron et al. \cite{baron2016weirdest}, where an unsupervised Random Forest was used to detect the most outlying galaxy spectra within the SDSS survey and its results were compared with with a standard Random Forest, a one-class Support Vector Machine and an Isolation Forest. The key difference is that in the present work the outliers are searched within images, instead of spectra, and including multiple kinds of objects in the dataset, instead of just galaxies.\\
The remainder of this work is organized as follows. We begin in Sec.~\ref{sec:exppre} with an exploration of the dataset used and its preparation procedure. Then, in  Sec.~\ref{sec:outlier} we describe and assess the performance of the various selected methods. The results are compared and analyzed in Sec.~\ref{sec:comp}, followed by a description of the model proposed to find outliers in new data in Sec.~\ref{sec:new}. Finally, we draw the conclusions and mention some perspectives for future work.

\section{The data}
\label{sec:exppre}

As template data set for our experiments we used a subset of data extracted from the Sloan Digital Sky Survey (SDSS, \cite{york2000sloan,sdss}) Data Release 9 (hereafter SDSS-DR9). 
This subset was originally assembled to estimate the photometric redshifts without any pre-classification in (cf. \cite{ refId0}). The extraction was performed randomly, trying to maintain the object class types as much as possible balanced in terms of their quantities, thus minimizing any risk of selection effect induced by the overabundance of a particular class.\\
It consists of $200,000$ stars, $200,000$ galaxies and $185,718$ quasars, represented by a OPS composed by the five bands (\textit{ugriz} filters).

\subsection{Data Exploration \& Preprocessing}
Prior to start the experiments, we performed an analysis of the extracted data, in order to identify  possible anomalies or artifacts that could affect the outlier detection performances.\\
A first consideration is that the spectroscopic and photometric classifications of the objects composing the SDSS dataset do not always match. For example, the SDSS object J145416.34+212953.9\footnote{\url{http://skyserver.sdss.org/dr12/en/tools/explore/summary.aspx?ra=223.568087062725\&dec=21.4983280618327}} is a galaxy according to the spectroscopy, but is listed as a star by the photometric classification. 
An example of the reverse is the SDSS object J170616.32+242609.2\footnote{\url{http://skyserver.sdss.org/dr12/en/tools/explore/summary.aspx?ra=256.568027662234\&dec=24.4358907062636}}, spectroscopically classified as a star, but resulting as a galaxy from the photometry. 
However, since this work is based on data for which the spectroscopy was always available, we  refer to the spectroscopic classification in all experiments.\\
Furthermore, a number of duplicate entries in the dataset were found: a couple of identical stars, two galaxy couples and $393$ couples of quasars. These duplicates were removed.\\

\begin{figure}
\begin{subfigure}{.5\textwidth}
  \centering
  \includegraphics[width=.9\linewidth]{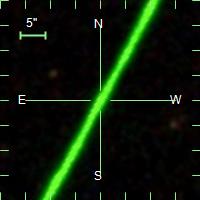}
  \caption{SDSS J161226.29+402805.3}

  \label{fig:gre}
\end{subfigure}%
\begin{subfigure}{.5\textwidth}
  \centering
  \includegraphics[width=.9\linewidth]{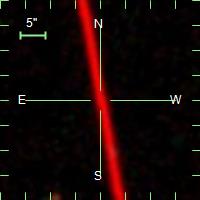}
  \caption{SDSS J171043.64+582219.5}
\label{fig:red}
\end{subfigure}
\caption{Examples of straight lines within the images, mostly due to Asteroids/satellites}
\label{fig:aster}
\end{figure}

Another example of irrelevant anomalies was the presence of strong straight lines passing through the r band, as for instance caused by a satellite or and asteroid. In order to remove such events from the dataset, a heuristic procedure was applied, by setting a magnitude a magnitude threshold in some band, able to reveal the occurrence of such events. 
By looking at all images where the maximum value in the r band was at least $1.75$ times higher than the highest value in all other bands, we found $106$ images affected by the straight line (considered as true positives), and $3$ images with a false anomaly event (called false positives). The true positives found with this criterion were removed from the data. An example is given in Fig.~\ref{fig:gre}.\\
By applying the same procedure on the other bands (u, g or i) with a threshold of $2$, an additional $88$ images were detected. Only four of these had a line present, all in the i band, similar to those found in the r band. Such a line was visible only for one object in the object explorer, see Fig.~\ref{fig:red}. These objects were not removed from the dataset.

Furthermore, we noticed that the five bands do not span the same range of pixel values, see Table~\ref{tab:meanstd}. Therefore, we decided to standardize all bands, using a scaling factor, able to obtain images with zero mean and unitary standard deviation \cite{kreyszig11}. In the cases where the dataset was split into a training and testing set, the standardization was performed using the mean and standard deviation of the training set as reference, in order to align both training and test distributions. 

\begin{figure}
\begin{subfigure}{.8\textwidth}
  \centering
  \includegraphics[width=\linewidth]{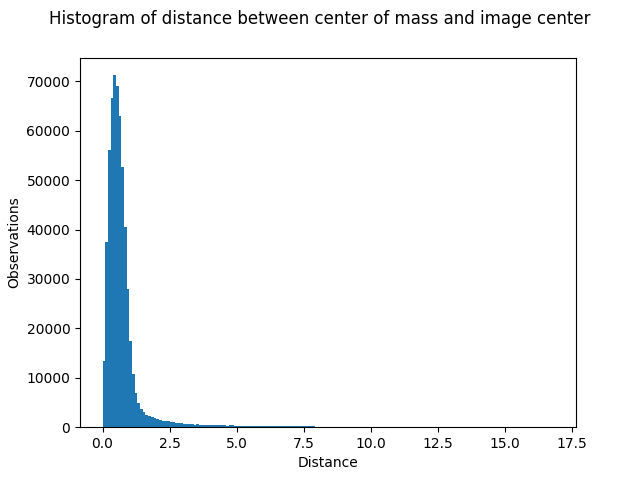}
  \caption{Whole dataset}
  \label{fig:com1}
\end{subfigure}%
\caption{Histogram of the distances between the image center of mass and image center}
\label{fig:com}
\end{figure}

\begin{table}
\centering
    \begin{tabular}{| c | c |c | }
    \hline
    \multirow{2}{*}{Filter} & \multirow{2}{*}{Mean}& Standard  \\ 
    & & Deviation \\\hline
     u  & 0.028 & 0.387\\ \hline
     g &  0.076 &0.769\\ \hline

     r & 0.133 & 1.290\\ \hline

     i & 0.179 & 1.628\\ \hline

     z & 0.224 & 2.183\\ \hline

    \end{tabular}
    \caption{Mean and standard deviations of the pixel values among the different bands}
    \label{tab:meanstd}
\end{table}
Finally, we were interested in detecting all occurrences of double sources. This was done by looking at the center of mass within the images, averaged on all bands. Fig.~\ref{fig:com} shows a histogram of the distances between the center of mass and the center of the images. There is  no obvious cut-off which could separate double sources and single sources.  Therefore, we performed a searching within SDSS archive, in order to find all candidate sources  to be paired. 
We used a cross-match threshold of $15$ pixels, corresponding to $0.6$ arcsec. We found $48,814$ galaxies, $26,126$ quasars and $33,130$ stars with at least one more object within this distance in our dataset. 

This information was not used to reject these sources from the dataset, but was taken into account in the post-processing phase in order to recognize this special category of objects.  

\section{Outlier Detection Methods}
\label{sec:outlier}
The outlier detection methods considered in this work can be broadly grouped in two categories: \textit{(i)} those applied on a dataset whose dimensionality is pre-reduced by Principal Component Analysis (Local Outlier Factor, Isolation Forest, K-Means, Modified Novelty Measure), and \textit{(ii)} deep learning models, for instance standard and convolutional autoencoders. \\

\subsection{PCA-based detection methods}
\begin{figure}
\begin{subfigure}{.9\textwidth}
  \centering
  \includegraphics[width=\linewidth]{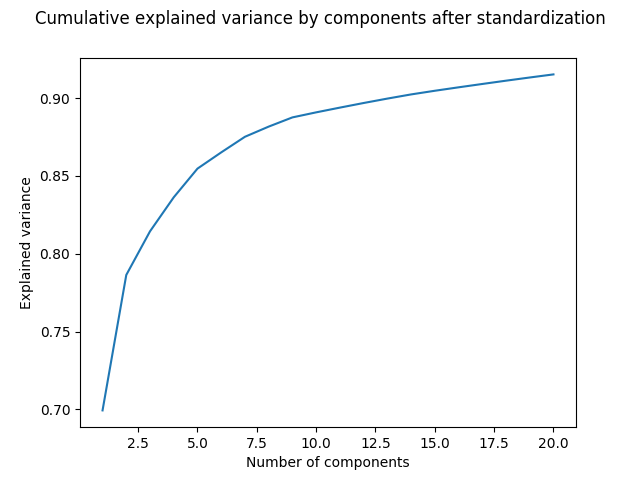}
\caption{After thresholding}
\label{fig:pca2}
\end{subfigure}%
\caption{Explained variance by number of PCA components, see Sec. \ref{sec:outlier}}
\label{fig:pca}
\end{figure}

Before  applying the outlier detection metrics on the pixel values, we decided to preliminarly reduce the dimensionality of the images.
 We used the sklearn implementation\footnote{\url{https://scikit-learn.org/stable/modules/generated/sklearn.decomposition.PCA.html}} of the widely used Principal Component Analysis (PCA) for this \cite{abdi2010principal}. \\
Two common approaches, for choosing the number of components to reduce the input data, are the elbow method \cite{Thorndike1953} and explaining some set amount of variance \cite{rea}.\\ 
As there is no clear elbow present in the variance graph of Fig.~\ref{fig:pca2}, we chose to explain a set amount, namely a 90\% of the variance, which comes down to $14$ components, see Fig.~\ref{fig:pca}.  

As stated in Sec. \ref{sec:outlier}, we used the features extracted by the PCA as input for a series of outlier detection methods, for instance the Local Outlier Factor, Isolation Forest, K-Means and Modified Novelty Measure, described in the following sections.

\subsubsection{Local Outlier Factor}
The Local Outlier Factor (LOF) is a measure for how isolated is an object with respect to its neighbourhood \cite{breunig2000lof}. 

\begin{figure}\centering
\begin{subfigure}{.8\textwidth}
  \centering
  \includegraphics[width=\linewidth]{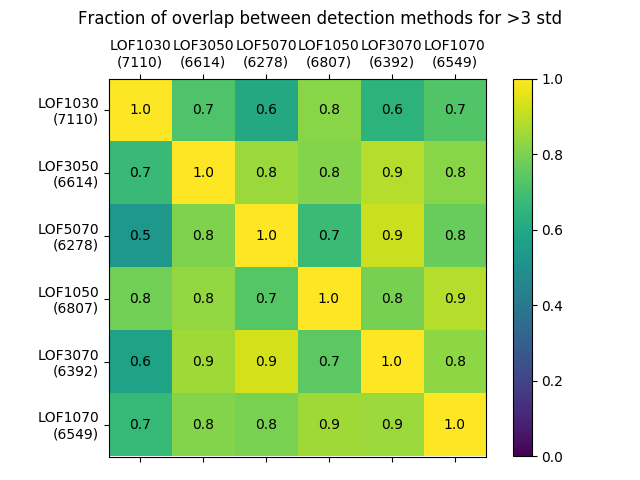}
  \caption{Between 3 and 5 standard deviations from the mean}
\end{subfigure}%
\\
\begin{subfigure}{.8\textwidth}
  \centering
  \includegraphics[width=\linewidth]{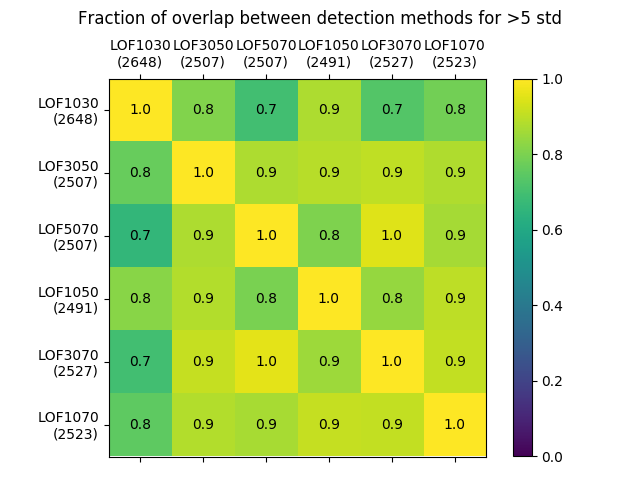}
\caption{Above 5 standard deviations from the mean}
\end{subfigure}
\caption{LOF results using different parameter settings. The name indicates the hyperparameter range used, e.g. LOF1030 used a range of [10, 30]. The size of the results is shown between brackets.}
\label{fig:lofmat}
\end{figure}

Let the k-distance $kd$ of an object be the distance to its k-th nearest neighbours, and let $N_k(p)$ be the set of those neighbours. The reachability distance of an object $p$ with respect to another object $o$, when considering $k$ nearest neighbours, is given by 
$$rd_k(p, o) = \max\{kd(o), d(p, o)\} ,$$
where $d(p, o)$ is the distance between points $p$ and $o$. Note that this is the actual distance between $p$ and $o$, if $p$ does not belong to the $k$ nearest neighbours of $o$, and the k-distance of $o$ if it does. Now define the local reachability density of $p$ considering $k$ nearest neighbours as 
$$ lrd_k(p) = 1 / \left( \frac{\sum_{o\in N_k(p)} rd_k(p, o)}{|N_k(p)|} \right) .$$
The LOF score for an object $p$ is then calculated by 
$$ LOF_k(p) = \frac{\sum_{o\in N_k(p)} lrd_k(o)}{lrd_k(p) |N_k(p)|} .$$
Intuitively, when the density around point $p$ is lower than the average of the densities around its $k$ neighbours, this fraction will result in a value above 1, in which case $p$ can be considered an outlier above some threshold.\\
The most important hyperparameter to set when calculating the LOF is the number of neighbours used. In their paper \cite{breunig2000lof}, the authors recommended taking a range. The lower bound should be seen as the minimum number of samples a cluster has to contain, such that other samples can be local outliers relative to this cluster. The upper bound can be instead considered the maximum number of close samples that can potentially be local outliers. The LOF value for each datapoint is computed for each value in this range and the maximum is taken as the final value.\\
The choice of the best bound in our case, by considering the reduced parameter space after having applied the  PCA, is not obvious. Therefore, we refer to the range introduced by the authors of the method, used in an analoguous case of a large dataset, namely $30-50$.\\
As the current scikit learn implementation\footnote{\url{https://scikit-learn.org/stable/modules/generated/sklearn.neighbors.LocalOutlierFactor.html\#sklearn.neighbors.LocalOutlierFactor}} does not support a range but only a scalar value for this hyperparameter, we computed the scores for the parameter values k = $30, 35, 40, 45, 50$ instead of at every integer value between $30$ and $50$ to save computing time. \\
The matrix shown in Fig.~\ref{fig:lofmat} provides infromation about the sensitivity to the hyperparameter setting.   Each cell in this matrix shows what fraction of objects  obtained by using the hyperparameter settings denoted in the column name are also present in results obtained by using the hyperparameter settings of the row name. For example, out of the 2,507 objects scoring above 5 standard deviations when using the range $[30, 50]$, approximately $80\%$ also score above $5$ standard deviations when using the range $[10, 30]$. This format is used throughout this work.\\
There is more overlap between the outliers scoring above 5 standard deviations from the mean, for the different hyperparameter ranges used, than for the outliers scoring between 3 and 5 standard deviations. All in all, apart from the [10, 30] range, the results are quite insensitive to the parameter settings, with the overlap mostly reaching over 80\% between 3 and 5 standard deviations and over 90\% above 5 standard deviations.

\subsubsection{Modified Novelty Measure}
\label{sec:nn}
A modified novelty (MN) measure originally designed for collider physics proposed, in Hajer et al. \cite{hajer2018novelty}, as an improvement to the LOF, is defined by
$$\frac{d^{-m}_{test} -d^{-m}_{train}}{d^{-m/2}_{train}} $$
where $d_{test}$ denotes the average distance from a point to its k nearest neighbours in the testing set, $d_{train}$ the average distance to its k nearest neighbours in the training set, and m the dimensionality.\\
In contrast to the LOF, which only considers one point at a time, this measure takes into account the clustering of testing data \cite{hajer2018novelty}. While we do not split our dataset for the calculation of the LOF, and as such testing data clustering is not an issue, it is still interesting to see how the results for the two measures differ. When applying the methods to novel data, this benefit will come into play.\\
The one hyperparameter to consider is the number of nearest neighbours used for the calculation of the metric. The authors do not provide guidelines on how to choose the hyperparameter for this method. As it serves a similar purpose to the hyperparameter used for LOF, we chose a value of 40, the average of the range of values used there. \\
There is no library currently implementing this method for Python. 
In contrast to the previous methods, this measure assigns a score very close to zero to its outliers, while the most normal points can get a very negative score. As a result it is not the outliers that score above 3 standard deviations but the most normal objects instead. \\


\subsubsection{Isolation Forest}
\begin{figure}  \centering

\begin{subfigure}{.8\textwidth}
  \centering
  \includegraphics[width=\linewidth]{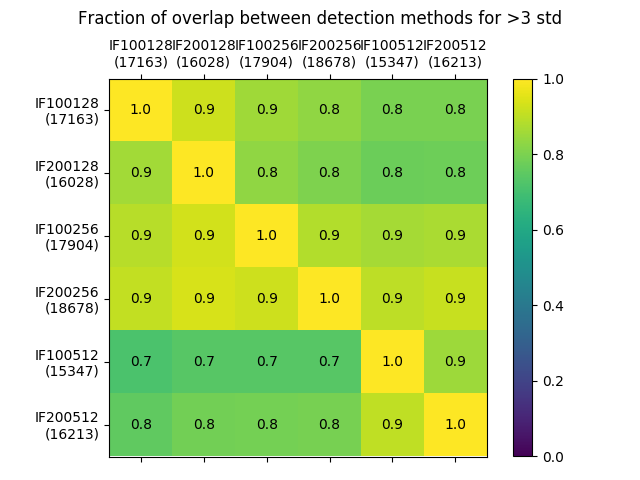}
  \caption{Between 3 and 5 standard deviations from the mean}
\end{subfigure}%
\\
\begin{subfigure}{.8\textwidth}
  \centering
  \includegraphics[width=\linewidth]{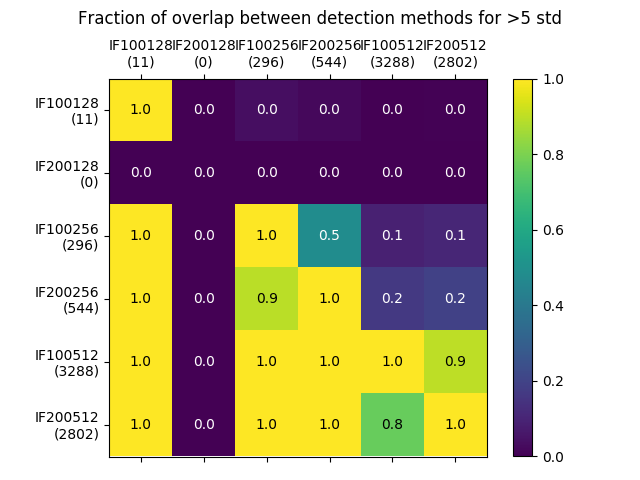}
\caption{Above 5 standard deviations from the mean}
\end{subfigure}
\caption{Overlap between the IF results using different parameter settings. The name indicates the hyperparameters used, e.g. IF100256 used 100 trees with 256 samples each. The size of the results is shown between brackets.}
\label{fig:ifmat}
\end{figure}

The Isolation Forest (IF, \cite{liu2008isolation}) method is based on the idea that an outlier will be separated from normal datapoints at an early stage when randomly partitioning the dataset. This is implemented by repeatedly partitioning a sampled subset of the dataset with a random value between the minimum and maximum value for a random feature until all data points are isolated, resulting in a tree structure. This process is repeated multiple times on different samples, creating a forest, and the mean path length to a data point over all trees in the forest is the outlier score for that point. \\
The two most important hyperparameters of IF are the number of trees in the forest, and the number of randomly sampled data points to use for the construction of each tree. In their paper the authors suggest using 100 trees, as the path lengths usually converge well before this point, as well as using a sub-sampling size of 256 for higher dimensional datasets \cite{liu2008isolation}.\\
In Fig.~\ref{fig:ifmat} we see the sensitivity of the detections to these hyperparameters. A logical consequence of using more samples is that the size of the set of objects with higher outlier scores becomes larger, as with bigger trees there will be larger differences between the average path lengths to isolation for a normal object and an outlier. For 100 trees with a sample size of 128 the consequence is that no object scores above 5 standard deviations and 17,163 between 3 and 5. On the other hand, with 100 trees of size 512 this is shifted to $3,288$ and $15,347$. This makes it more difficult to compare the sets with different sample sizes, especially above 5 standard deviations. \\
Between 3 and 5 standard deviations the results are very similar for sample sizes of 128 and 256. For a sample size of 512 there is less overlap. This can be explained by the fact that, due to the larger trees, the outliers scoring between 3 and 5 standard deviations, when using the lower sample sizes, are now scoring too high to be included in this set and are now scoring above 5 standard deviations. Note that the trees are not built from the same random samples, so the small differences in outliers can be caused by which data points were used to build the trees, instead of what hyperparameter settings were chosen.

\subsubsection{Clustering}
\begin{figure}
\centering
      \includegraphics[trim=0 1cm 0 0, clip, width=0.6\linewidth]{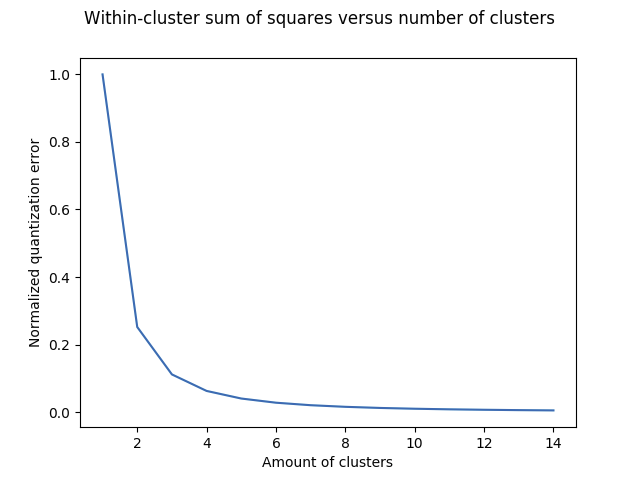}\\
    Number of Clusters
    \caption{Normalized quantization error by number of clusters used}
    \label{fig:quant}
\end{figure}

\begin{figure}
\begin{subfigure}{.33\textwidth}
  \centering
  \includegraphics[width=\linewidth]{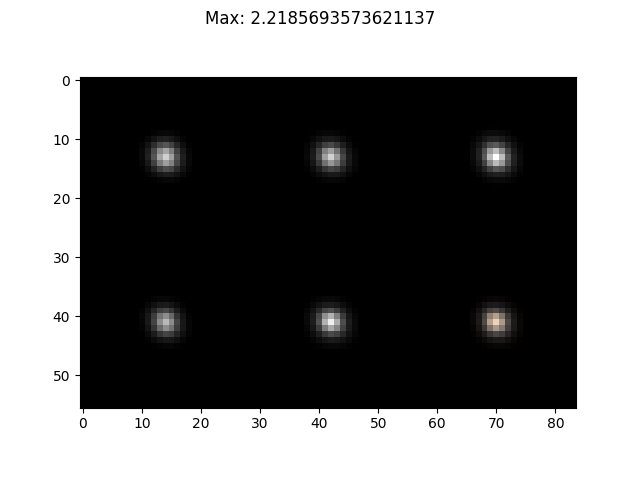}
  \caption{First cluster prototype}
\end{subfigure}%
\begin{subfigure}{.33\textwidth}
  \centering
  \includegraphics[width=\linewidth]{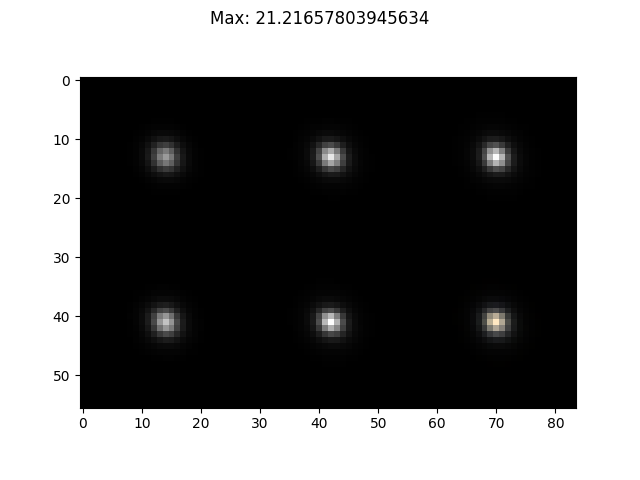}
  \caption{Second cluster prototype}
\end{subfigure}%
\begin{subfigure}{.33\textwidth}
  \centering
  \includegraphics[width=\linewidth]{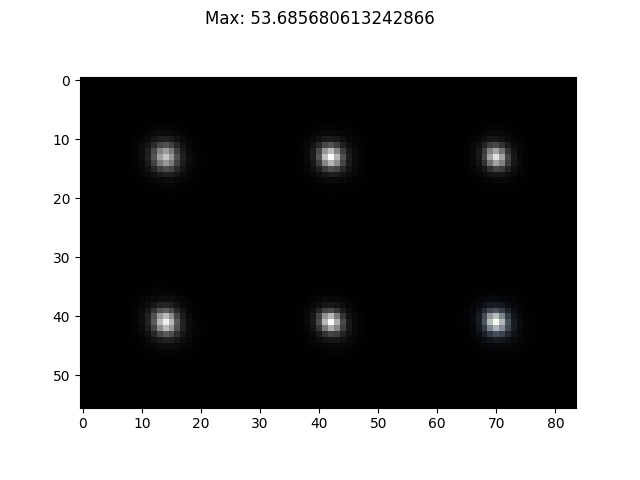}
  \caption{Third cluster prototype}
\end{subfigure}
\caption{Cluster prototypes when using k = 3 and Euclidean distance}
\label{fig:proto}
\end{figure}

Another approach to detecting outliers is by grouping the dataset into clusters and using the distance between objects and their closest cluster center as the outlier score \cite{wei}. One of the most commonly used clustering algorithms is k-means (KM,  \cite{lloyd1982least}). The only hyperparameter of this algorithm determines how many cluster prototypes will be generated. Points are then assigned to the closest prototype, after which the location of the prototypes is recomputed as the mean of all points assigned to it. This is repeated until convergence \cite{lloyd1982least}.\\
As there are three types of objects in our dataset the logical choice for k is three. We can verify this choice by plotting the quantization error against the number of clusters. At a value of three there is a slight kink in the line, suggesting that this would be a good choice for k (Fig.~\ref{fig:quant}). We noticed that the absence of a true elbow in the curve gives a first indication that our data is not partitioned into well-defined clusters using KM.\\
The clusters are not equally populated, shown in Table~\ref{tab:numclusts}, and each cluster does not represent each type of object. Visualizing the prototypes by reverting the PCA gives the results shown in Fig.~\ref{fig:proto}. The defining difference between the clusters seems to be the maximum value. This explains the uneven cluster membership as the prototype with the lowest value, the second cluster, has by far the largest number of objects associated with it, and by far the largest number of objects in our dataset have a low maximum value (Fig.~\ref{fig:maxHists2}).

\begin{table}
    \centering
    \begin{tabular}{| l | l | }
    \hline
    - & Members \\ \hline
     0  & 47,883 \\ \hline
     1 &  517,110\\ \hline

     2 & 13,097 \\ \hline

    \end{tabular}
\caption{Cluster membership with k = 3}
    \label{tab:numclusts}

\end{table}

\begin{figure}  \centering

\begin{subfigure}{.8\textwidth}
  \centering
  \includegraphics[width=\linewidth]{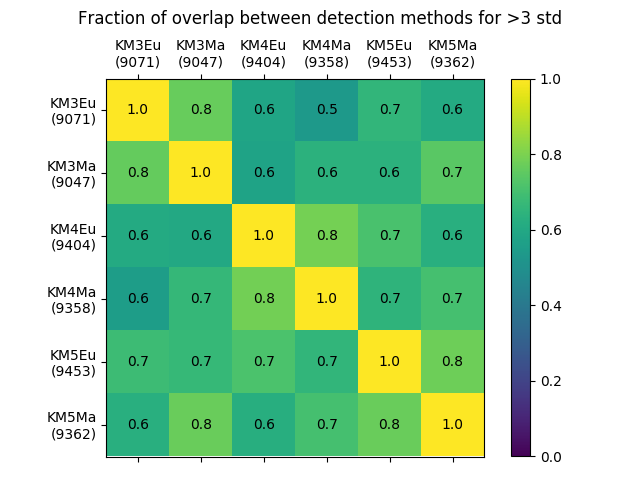}
  \caption{Between 3 and 5 standard deviations from the mean}
\end{subfigure}%
\\
\begin{subfigure}{.8\textwidth}
  \centering
  \includegraphics[width=\linewidth]{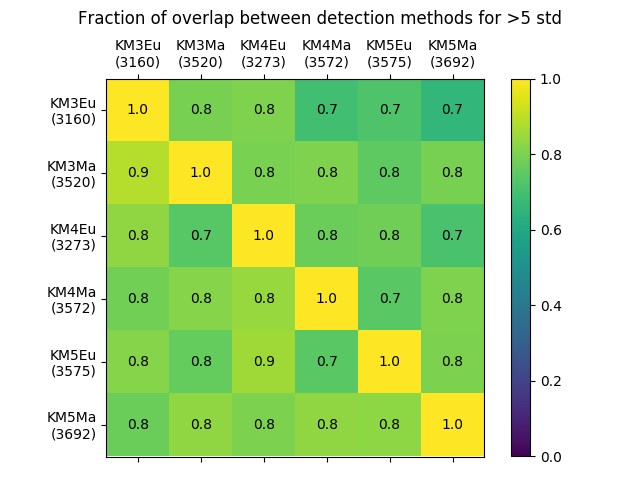}
\caption{Above 5 standard deviations from the mean}
\end{subfigure}
\caption{Overlap between the KM results using different parameter settings. The name indicates the hyperparameters used, e.g. KM3Eu used 3 clusters and measured the Euclidean distance. The size of the results is shown between brackets.}
\label{fig:kmmat}
\end{figure}

\begin{figure}
    \centering
    \includegraphics[width=0.8\linewidth]{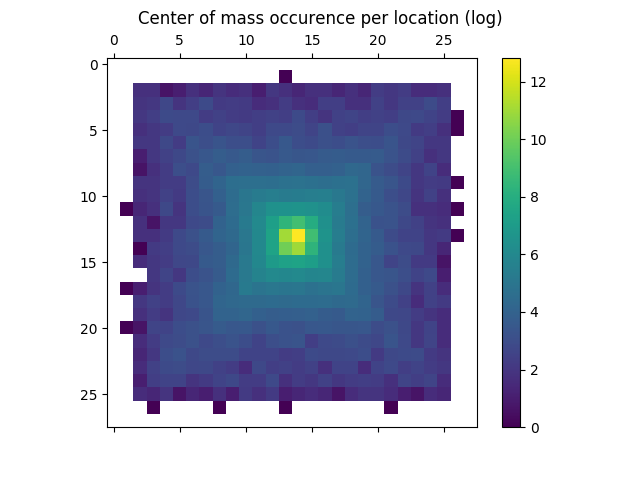}
    \caption{Image pixels colored by center of mass occurrence}
    \label{fig:comimg}
\end{figure}

The distance metric used when determining the outliers after clustering is another hyperparameter. A standard choice is the $L_k$ norm for some value k, defined by
$$L_k{(x, y)} = \sum_{i = 1}^d (\norm{ x^i - y ^i}^k)^{1/k}$$
When using the $L_k$ norm for KM with high-dimensional data, often a lower k is preferred \cite{aggarwal2001surprising}. As the current scipy implementation only supports integer values for k\footnote{\url{https://docs.scipy.org/doc/scipy/reference/spatial.distance.html}}, we use the Manhattan distance (k = 1). \\
We can still gain insights into the sensitivity of the results to the choice of these hyperparameters by looking at the overlap matrix, found in Fig.~\ref{fig:kmmat}. Especially the lower scoring outliers are different when using a different value for k.  Where the maximal values of the prototypes for $k = 3$ have rounded values of 2, 21 and 54, for $k = 4$ we have 2, 14, 37 and 63, and for $k = 5$ they are 1, 9, 25, 49 and 74, the morphologies for the prototypes with higher k are very similar to the ones in Fig.~\ref{fig:proto}. Apparently the KM is producing a clustering mostly based on the maximum value of the pixels, rather than a proper partition.
They are unable to take another shape as the center of mass is almost uniformly distributed around the center, as illustrated in Fig.~\ref{fig:comimg}. A convolutional approach might be favorable to deal with this problem.

{}

\subsection{Autoencoders}
Autoencoders are a fundamental architecture for unsupervised outlier detection using deep learning \cite{chalapathy2019deep}. Autoencoders consist of two mirrored parts, an encoder and a decoder. The idea is that the encoder tries to learn a representation of the input in a lower dimensionality, while the decoder rebuilds the original input from the smaller representation. Ideally the autoencoder as a whole behaves the same as the identity function.\\
Autoencoders can be used in two ways to detect outliers. The first way is to feed an input to the trained model and look at the difference between the output and the input. This difference is called the reconstruction error. The bigger this error is, the more outlying the input is with respect to the data the autoencoder was trained with. Secondly, one can use the lower dimensional representation obtained after the encoder as a new feature vector to apply other outlier detection methods on. We use the first method in this work.

\subsubsection{Standard Autoencoder}
The simplest autoencoders (AE, \cite{lyudchik2016outlier}) treat the input as a one-dimensional feature vector with in our case $28*28*5=3920$ elements (five images, one for each band of 28x28 pixels). As we are dealing with image-like data, and tuning and training autoencoders is a time-consuming process, we focused on tuning the convolutional autoencoder as it should be better suited to our dataset. For this reason the normal autoencoder is included in the visual comparisons in Sec.~\ref{sec:comp}, to get an idea if the results are different from other variants, but the results are not used in the final comparisons.\\
The model we used for the comparison reduces the 3,920 dimensional input to 128 dimensions with 3 hidden layers for both the encoder and decoder part, see Table~\ref{tab:aearch} for the details. 
\begin{table}
    \centering
    \begin{tabular}{| l | l | l | l | l | l | l | l |}
    \hline
    Layer & 0 & 1 & 2 & 3 & 4 & 5 & 6 \\ \hline
     Number of neurons  & 3,920 & 784 & 256 & 128 & 256 & 784 & 3,920 \\ \hline
    \end{tabular}
\caption{Standard autoencoder architecture}
    \label{tab:aearch}

\end{table}

\subsubsection{Convolutional Autoencoder}
Convolutional autoencoders (CAE) do not treat each individual measurement as a separate entity, but instead use small filters to extract and detect local image features \cite{ribeiro2018study}. \\
The performance of neural networks such as autoencoders can be very sensitive to the hyperparameters used. Unlike the previous PCA-based methods, the training of the CAE is a supervised procedure, enabling the use of hyperparameter optimization algorithms such as random search \cite{bergstra2012random}.
We used the hyperparameters of the best performing network obtained from the random search to train on the whole dataset and used this for the comparison in the next chapter. The input is reduced by 2 hidden layers to a 7x7 image. Each hidden layer consists of a convolutional layer with a filter size of 3x3, followed by a max pooling layer for the encoder and an upsampling layer for the decoder, both of size 2x2. A learning rate of 0.075 was used, linearly decaying to 0 over the epochs, finally we used a stride of 1x1.

\begin{table}
    \centering
    \begin{tabular}{| l | l | l | l | l | l | l | l |}
    \hline
    Layer & 0 & 1 & 2 & 3  \\ \hline
     Number of convolutions  & 48 & 24 & 24 & 48 \\ \hline
    \end{tabular}
\caption{CAE architecture}
    \label{tab:caearch}

\end{table}

\begin{figure}  \centering

\begin{subfigure}{.8\textwidth}
  \centering
  \includegraphics[width=\linewidth]{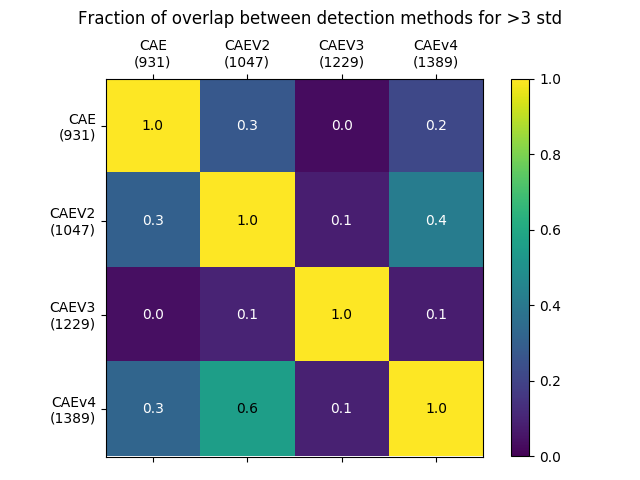}
  \caption{Between 3 and 5 standard deviations from the mean}
\end{subfigure}%
\\
\begin{subfigure}{.8\textwidth}
  \centering
  \includegraphics[width=\linewidth]{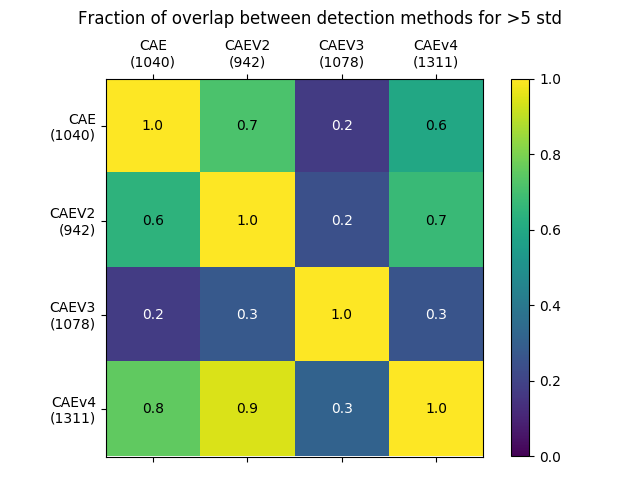}
\caption{Above 5 standard deviations from the mean}
\end{subfigure}
\caption{Overlap between the CAE results using different parameter settings. The size of the results is shown between brackets.}
\label{fig:CAEmmat}
\end{figure}

In Fig.~\ref{fig:CAEmmat} we look at the overlap in outliers when training a CAE with different values for the number of hidden layers, the number of convolutions in each layer, the learning rate and the batch size. The first 3 versions are trained with hyperparameters tested in the random search. The 4th version uses the same settings as version 2 but is trained on the whole dataset and is the one used for our comparisons. 
As the CAEs were trained using a different training/testing split, at least part of the difference will be because of testing data being inherently more likely to have a higher reconstruction error than the training data.\\
The third iteration had a total reconstruction error more than 6 times higher than the first two ($\pm 0.033$ versus $\pm 0.20$). Its outliers are also very different from the others, never reaching aan overlap above 30\%. While the first two had a similar total reconstruction error, the outliers differ, especially between 3 and 5 standard deviations, but it is difficult to determine how much of this can be attributed to the training/testing split. 

\section{Discussion}
\label{sec:comp}

\begin{figure}
\begin{subfigure}{.5\textwidth}
  \centering
  \includegraphics[width=.9\linewidth]{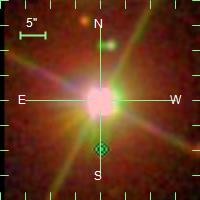}
  \caption{SDSS J053838.82+053809.2}
  \label{fig:1481}
\end{subfigure}%
\begin{subfigure}{.5\textwidth}
  \centering
  \includegraphics[width=.9\linewidth]{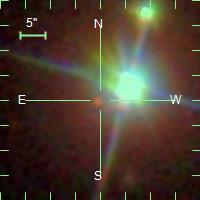}
  \caption{SDSS J090331.62+220120.0}
\label{fig:1482}
\end{subfigure}
\caption{Two examples of images showing the highest values in terms of pixel's intensity among all SDSS bands. It is possible to see a set of sources close to a very bright star nearby}
\label{fig:148}
\end{figure}

\begin{figure}
\begin{subfigure}{.5\textwidth}
  \centering
  \includegraphics[width=\linewidth]{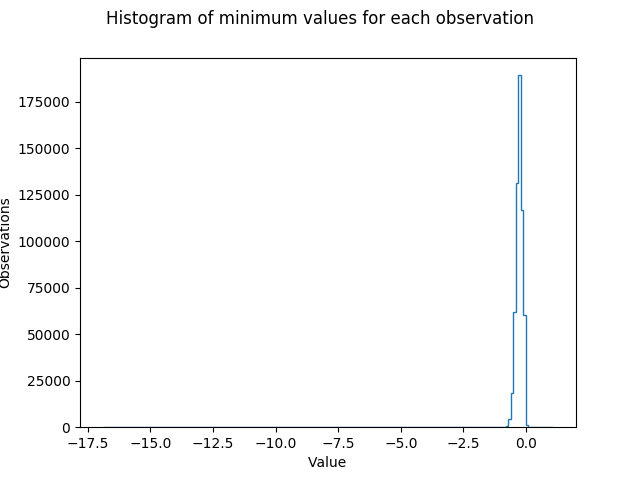}
  \caption{Minimum values}
  \label{fig:maxHists1}
\end{subfigure}%
\begin{subfigure}{.5\textwidth}
  \centering
  \includegraphics[width=\linewidth]{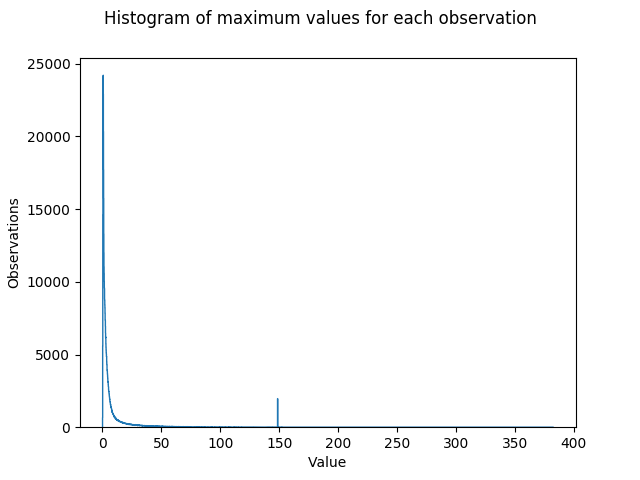}
  \caption{Maximum values}
\label{fig:maxHists2}
\end{subfigure}
\label{fig:maxHists}
\caption{Histogram the distribution of pixels with the maximum value among all SDSS bands.}
\end{figure}

Through an exploration of the input dataset, composed by the image cutouts extracted from the SDSS, an analysis at pixel level of images was performed. By projecting on a histogram the distribution of pixels with the maximum value among all involved bands (shown in Fig.~\ref{fig:maxHists2}), it appeared a peak at $\sim148$. Looking at the sources around the peak, these were recognized as objects located nearby to a very bright star. Two examples of such objects are given in Fig.~\ref{fig:148}. 

When we look at the maximum value of pixel for each object (over all bands), we get the histogram shown in Fig.~\ref{fig:maxHists2}. 

\begin{figure}  \centering

\begin{subfigure}{.8\textwidth}
  \centering
  \includegraphics[width=\linewidth]{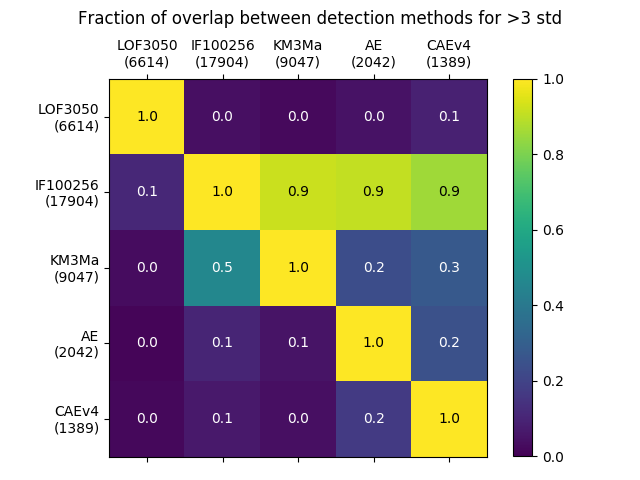}
  \caption{Between 3 and 5 standard deviations from the mean}
\end{subfigure}%
\\
\begin{subfigure}{.8\textwidth}
  \centering
  \includegraphics[width=\linewidth]{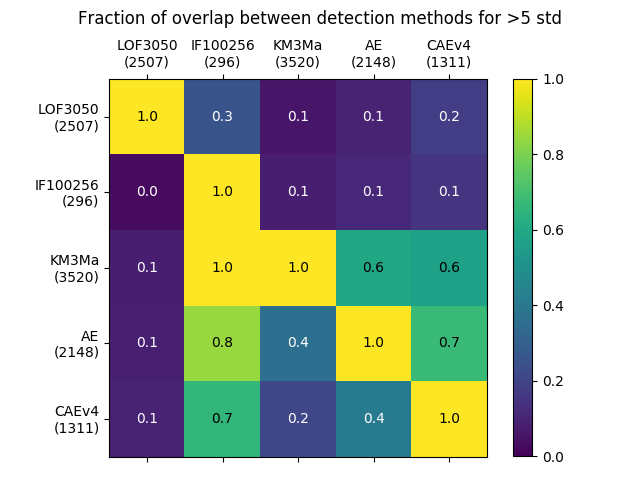}
\caption{Above 5 standard deviations from the mean}
\end{subfigure}
\caption{Overlap between the different outlier detection methods. The size of the results is shown within brackets.}
\label{fig:difmat}
\end{figure}

Since such kind of objects were affected by the presence of the much brighter second source, they were considered as unreliable and removed from the dataset.
We decided to cut the images with a maximal value higher than $100$ from the dataset. This threshold was chosen because, when running the outlier detection methods on the whole dataset, only five outliers were detected with a maximum value below $50$, none with a maximum value between $50$ and $100$ and the rest with a maximum value over $100$. Of the latter, all objects are child sources close to another, very bright object. Applying this threshold we removed $300$ galaxies, $97$ quasars and $7,232$ stars, leaving a total of $578,090$ objects. The histogram of minimal values shown in Fig.~\ref{fig:maxHists1} does not show any unexpected peak.\\
In terms of outlier detection, we decided to compare the results from different methods, by considering the overlap of objects scoring above $5$ standard deviations from the mean and between $3$ and $5$ standard deviations from the mean for the different outlier detection methods.
This is shown in Fig.~\ref{fig:difmat}. Note that the modified novelty measure is not included in this comparison here, due to the absence of outliers scoring above $3$ standard deviations for this method.

\begin{figure}
\begin{subfigure}{.5\textwidth}
  \centering
  \includegraphics[width=.9\linewidth]{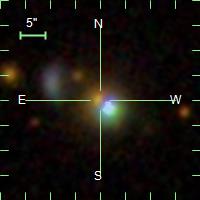}
  \caption{SDSS J081328.10+205556.6}
  \label{fig:sfig3}
\end{subfigure}%
\begin{subfigure}{.5\textwidth}
  \centering
  \includegraphics[width=.9\linewidth]{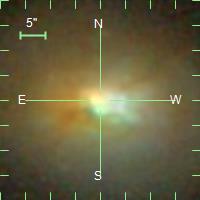}
  \caption{SDSS J020938.56-100846.1}
  \label{fig:sfig4}
\end{subfigure}
\begin{subfigure}{.5\textwidth}
  \centering
  \includegraphics[width=.9\linewidth]{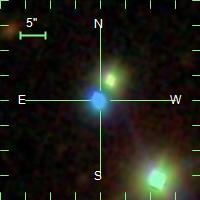}
  \caption{SDSS J065437.72+274200.4}
  \label{fig:sfig5}
\end{subfigure}%
\begin{subfigure}{.5\textwidth}
  \centering
  \includegraphics[width=.9\linewidth]{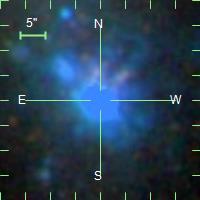}
  \caption{SDSS J125900.31+345042.8}
  \label{fig:weirde}
\end{subfigure}
\caption{4 objects scoring above 5 standard deviations for LOF, IF, KM and CAE}
\end{figure}

We explored the possibility to visualize the results by reducing the dimensionality with two different methods, PCA and t-SNE \cite{maaten2008visualizing} but without noticing any clear clusters.
What we saw from both the visualization and the correlation matrices is that the LOF gives very different results than the other methods. The other PCA-based methods and the normal autoencoder produce more similar results, especially when looking at the visualization. The CAE, while not as different as the LOF, does have less overlap with the other methods.\\
We found $63$ objects that score above $5$ standard deviations from the combination of LOF, IF, KM and CAE methods. 
They consist of $6$ galaxies, no quasars and $57$ stars. Of the $57$ stars, $35$ are observations with $2$ bright stars present in the frame. While these are definitely outliers from a data perspective, as they have two relatively bright objects, they are not the most interesting from an astronomical point of view. There are $4$ detections with triple stars that, although rarer, are equally uninteresting. The rest are either big, blue single stars or objects labeled as galaxies very close to bright stars. The $6$ galaxies vary from the reflection of a bright star in the telescope to irregular starburst regions, but do not share common traits. Four of the weirdest objects can be found in Fig.~\ref{fig:weirde}. 


Out of the $63$ objects, $12$ have no other sources in the image within $0.6$ arcsec from the object center, respectively, $9$ stars and $3$ galaxies. One galaxy is an artifact caused by a reflection in the telescope, while the other $2$ are either close by or interacting with a nearby galaxy further away than the threshold. Out of the $9$ stars, for $2$ of them the bands are moved, $1$ is close to an artifact, while the other $6$ are relatively big, blue stars.\\
When looking at objects that score over $5$ standard deviations for some methods and between $3$ and $5$ for others, the number of objects found does not increase as expected. This is again caused by the LOF way of finding mostly different outliers, compared to the other $3$ methods, as shown in Table~\ref{tab:inter}. None of the filtered asteroids showed up in the intersections of the combination of the $4$ methods.
\begin{table}
    \centering
    \begin{tabular}{| l | c | c | c | c | c |}
    \hline
    number of methods with $score>5\sigma$ & 4 & 3 & 2 & 1 & 0 \\ 
    number of methods with $5>=score>3\sigma$ & 0 & 1 & 2 & 3 & 0 \\ \hline
    methods & \multicolumn{5}{c|}{number of objects} \\ \hline
    LOF / IF / KM / CAE  & 63 (12) & 134 (39) & 153 (35) & 112 (24) & 44 (6)\\ \hline
    IF / KM / CAE  & - & 194 & 593 & 890 & 382 \\ \hline
    LOF / IF / KM  & - & 75 & 195 & 350 & 225 \\ \hline
    \end{tabular}
\caption{Number of objects that score above $5$ for some models and between $3$ and $5$ standard deviations for other models when considering those given in the row name. For example, there are 153 objects that, when taking into account all four models, score above 5 standard deviations for two of them, and between 3 and 5 standard deviations for the others. The number of single source objects is reported between brackets.}
    \label{tab:inter}
\end{table}

\begin{figure}\centering
\begin{subfigure}{.8\textwidth}
  \centering
  \includegraphics[width=\linewidth]{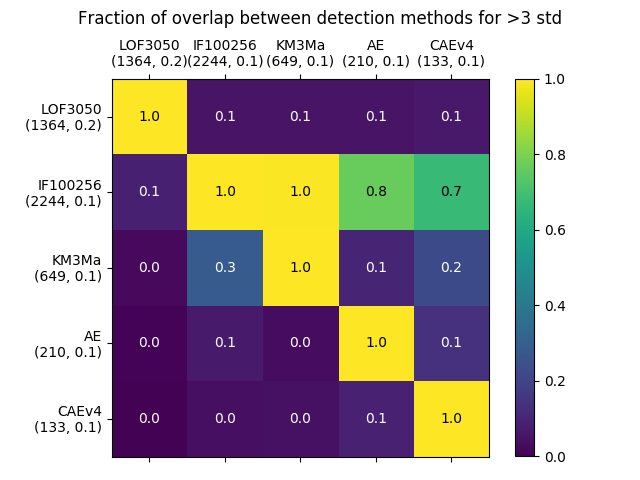}
  \caption{Between 3 and 5 standard deviations from the mean}
\end{subfigure}%
\\
\begin{subfigure}{.8\textwidth}
  \centering
  \includegraphics[width=\linewidth]{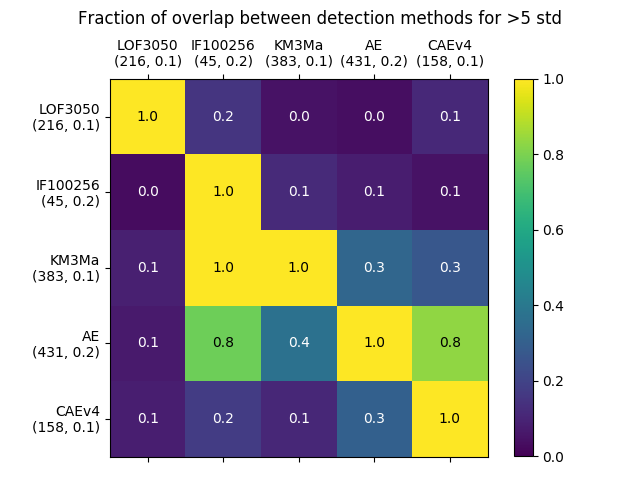}
\caption{Above 5 standard deviations from the mean}
\end{subfigure}
\caption{Overlap between the different outlier detection methods on galaxies. The size of the results and the fraction of the total results this constitutes is shown within brackets.}
\label{fig:difmatgal}
\end{figure}

\begin{figure}  \centering

\begin{subfigure}{.8\textwidth}
  \centering
  \includegraphics[width=\linewidth]{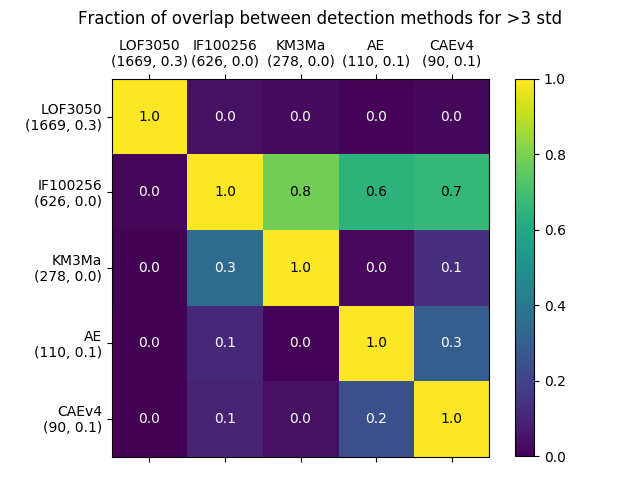}
  \caption{Between 3 and 5 standard deviations from the mean}
\end{subfigure}%
\\
\begin{subfigure}{.8\textwidth}
  \centering
  \includegraphics[width=\linewidth]{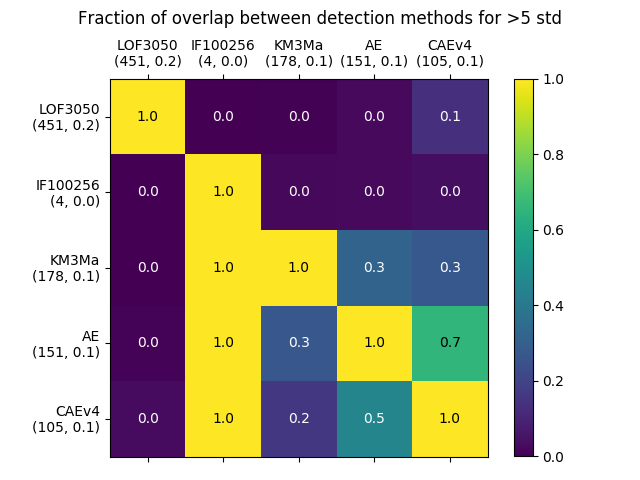}
\caption{Above 5 standard deviations from the mean}
\end{subfigure}
\caption{Overlap between the different outlier detection methods on quasars. The size of the results and the fraction of the total results this constitutes is shown within brackets.}
\label{fig:difmatqua}
\end{figure}

\begin{figure}  \centering

\begin{subfigure}{.8\textwidth}
  \centering
  \includegraphics[width=\linewidth]{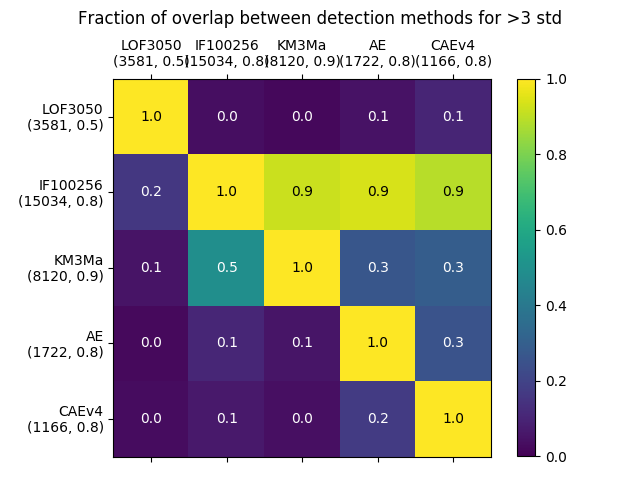}
  \caption{Between 3 and 5 standard deviations from the mean}
\end{subfigure}%
\\
\begin{subfigure}{.8\textwidth}
  \centering
  \includegraphics[width=\linewidth]{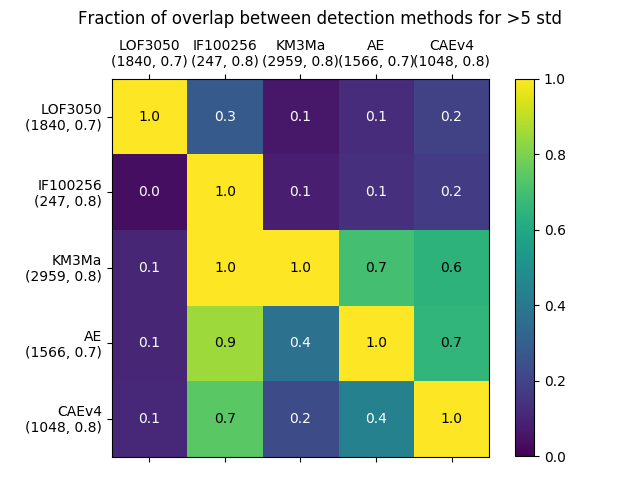}
\caption{Above 5 standard deviations from the mean}
\end{subfigure}
\caption{Overlap between the different outlier detection methods on stars. The size of the results and the fraction of the total results this constitutes is shown between brackets.}
\label{fig:difmatstar}
\end{figure}

\begin{figure}
  \centering
  \includegraphics[width=.75\linewidth]{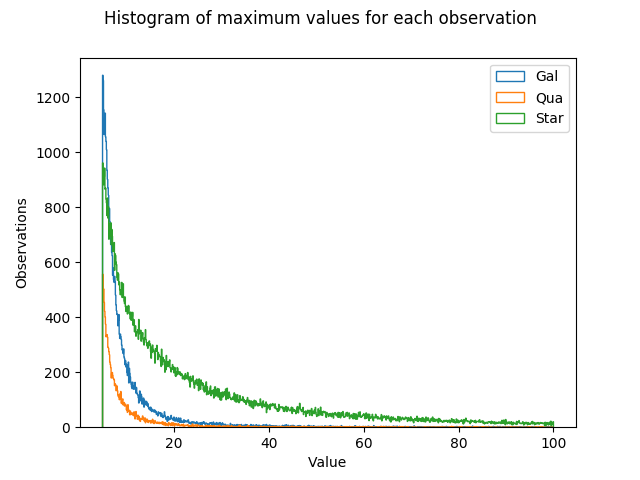}
  \caption{Histogram of the maximal value of pixels, among all SDSS bands, for Galaxies, Stars and QSOs. }
  \label{fig:maxhistcut}
\end{figure}

Splitting the results into the three classes of objects (Stars, Galaxies and QSOs) present and looking at their overlaps, could give us an idea if there is one method that filtered the outlying galaxies and quasars out of the intersection. The correlation matrices are found in Fig.~\ref{fig:difmatgal} for galaxies, Fig.~\ref{fig:difmatqua} for quasars and Fig.~\ref{fig:difmatstar} for stars. The CAE outliers consist of around 80\% stars in both categories, followed by around 10\% galaxies and slightly less quasars. The other methods except LOF follow the same ordering. The LOF finds more quasars than galaxies, and both categories are better represented in the results, with stars only making up around 50\% of the objects scoring between 3 and 5 standard deviations.\\
The fact that all these stars are detected as outliers again indicates that an high  maximum value of pixels, among all SDSS bands,  make for strong outliers. In Fig.~\ref{fig:maxhistcut} we see that stars are the brightest class present in the dataset, followed by galaxies, while quasars have the lowest intensity. This is in line with the objects with high outlier scores for all methods except LOF.\\
Similar to using the SDSS database to find double objects, this split into the object classes cannot be performed on new datasets and as such is only used to understand the results.

By looking at the strongest scoring objects per method we can get a general idea of what kind of objects the methods consider especially outlying. \\
The highest scoring objects for the LOF are the most varied. The top 250 objects include 19 galaxies and 50 quasars. The lowest maximal image value is 0.15 and the highest 94.33, and there are 108 single sources. There are 24 asteroids in the top 250 that are filtered from the results as discussed in Sec.~\ref{sec:exppre}, while none of the others have a single one of these in their 250 highest scoring objects. There are 8 objects with an entirely flat band in this list, whereas none of the others score such an object this highly.\\
The strongest IF outliers are mostly very bright objects. The lowest maximal intensity is 4.10, which is an entirely blue field due the reflection of a star, but where the LOF has 27 objects with a maximal value below 1, the IF has 27 objects below 50. There are 10 objects with a maximum value of exactly 50.93, all big blue stars. The IF is the strongest filter of quasars, with only 3 quasars scoring in the top 250. \\
The KM results on the other hand contain the most single sources, 187 out of 250. It also has the largest number of stars, namely 210. A large number of these are big, blue stars instead of the double sources we see more often for the other methods. 15 of these have a maximal value of exactly 50.93, of the in total 17 of these that exist in our dataset. While not as extreme as IF, the KM results are also mostly very bright objects. with only 3 of them having a maximal intensity below 25.\\
Finally, for the CAE we again have a more diverse list with 4 objects having a maximal value below 1 and the different types of objects better represented, even though not as varied as the LOF results. While the other 3 methods do not have a single object scoring in the top 250 with a center of mass over 8 pixels away from the image center, there are 58 objects that do so for the CAE. These are all relatively faint objects with a very bright star at the image edges. This method thus seems to consider double sources as the most outlying.\\
While for the method detecting the least single sources still over half of the strongest 250 objects contain a single source, the intersection between the methods contains comparatively a very small number of single sources, see Table~\ref{tab:inter}. \\
Taking the above observations into consideration, we can look at the objects scoring above 5 standard deviations for 3 methods and between 3 and 5 for one method to see if they match our expectations. 
We found that 86 out of the 135 objects have a lower score for IF, most of which will be caused by the small number of objects scoring above 5 standard deviations for IF. There are no objects only scoring between 3 and 5 standard deviations for KM. The 7 objects that do so for the CAE seem representative of the objects scoring high for all methods and do not share a common trait. The 42 objects scoring lower for the LOF all have a high intensity, the lowest being 50.93, which is in line with the expectations.\\
We tried a number of ideas, such as normalizing each object independently, but the resulting outliers are simply noisy images as these tend to have strange intensity distributions.\\
Secondly, as we know that each image is centered on the object, we tried segmenting this object and using it as a mask on the original observation. The distance between masked background and the lowest intensity value within the object, as well as the degree to which the intra-object values are scaled, can be considered hyperparameters that influence the trade-off in importance between morphology and intensity. Segmenting the object with a threshold, followed by morphological closing, however, makes too many assumptions about the underlying object. \\

Therefore, taking into account the four methods used in our comparison, the strongest outliers are dominated by images with multiple, relatively bright sources. Clearly, some pre-processing step is needed to get rid of multiple sources as well as enabling low intensity objects to reach a higher score. Deciding which method outperforms the others, if any, is not possible without a domain expert, analyzing the most outlying results one by one. We can conclude that the CAE and especially the LOF find mostly different outliers than the other more similar methods discussed in this work, and that taking the intersection of the highest scoring objects for each method does result in a list of manageable size for further inspection.\\
In conclusion, despite at this stage we did not find very interesting objects, the proposed strategy is actually capable to identify outliers, like satellites, asteroids or problematic images and we think that, when applied to data of a higher quality, for instance with a deeper and higher quality photometry, possibly extending to IR bands, it could lead to more interesting results. \\

\subsection{Future work}
\label{sec:new}

There is a variety of possible improvements for the methods applied in this work.\\ 
The first could be, instead of using PCA for the dimensionality reduction, to use more involved options, such as the features extracted by a CNN or the lower dimensional representation of the CAE. They could be used as the input to the outlier detection methods.\\
Using all integers for the LOF as hyperparameter values in the chosen range, instead of using intervals of 5, is in line with the recommendation of the authors. This can efficiently be computed by calculating the required values for the upper bound number of neighbours, and using these pre-computed values to quickly calculate the LOF values in the whole range \cite{breunig2000lof}.\\ 
As for the IF, a variant, such as the Extended Isolation Forest (EIF), can be used. The EIF claims to increase the reliability and consistency of the outlier scores compared to those of vanilla IF, which can produce artifacts as its decision boundaries are either vertical or horizontal \cite{hariri2018extended}. When applied to spectra, IF is unable to find more subtle outliers, which are often more interesting from an astronomical standpoint. Restricting the value at which to split at a certain node, to be for example between the 10th and 90th percentile of the feature value, improved the performance \cite{baron2019machine}.\\ 
Variants of KM such as fuzzy k-means can be tried out, as well as many other different clustering algorithms.\\
For the CAE, further testing with a wider range of hyperparameters and their values can be performed to improve its performance. Whether this also improves the quality of the outliers found can also be investigated.\\  
Creating the mask of the center object could be done by more sophisticated methods than thresholding, such as using a object detection and segmentation tool like MTObjects to detect the shape of the object \cite{teeninga2016statistical}. These could also be used to detect observations with multiple sources present without the need for a database.\\
For the LOF, KM and MN methods, a decision about which distance metric to use has to be made. The effect of different values for k in the $L_k$ norm or using a different distance metric altogether, such as the cosine distance, remains to be investigated.\\
Finally, an adaptation of the unsupervised random forest algorithm, which outperformed IF and some other outlier detection methods on the SDSS galaxy spectra, can be tried out on our dataset \cite{baron2016weirdest}.

\bibliographystyle{spphys}
\bibliography{main}{}

\begin{acknowledgement}
The authors wish to thank Prof. Michael Biehl for useful discussion and contribution which improved the quality of the work.
All tables and graphs were made using Matplotlib \cite{Hunter:2007}. The programs themselves make heavy use of the Scipy library \cite{scipy}. The autoencoders are made using Keras \cite{chollet2015keras}. SC acknowledges the financial contribution from FFABR 2017. MBr acknowledges financial contributions from the agreement \textit{ASI/INAF 2018-23-HH.0, Euclid ESA mission - Phase D} and the \textit{INAF PRIN-SKA 2017 program 1.05.01.88.04}.
\end{acknowledgement}

\end{document}